\documentclass[12pt, oneside, reqno]{amsart}
 \usepackage{graphicx}
\usepackage{amsmath,amsthm,amsfonts,amscd,amssymb,mathrsfs,hyperref}
\usepackage[all]{xypic}
\usepackage{amscd}   
\usepackage[all]{xy} 
\usepackage{booktabs} 
\usepackage{physics}
\usepackage{hyperref}
\hypersetup{colorlinks=true,citecolor=blue, urlcolor= cyan}
\usepackage{qcircuit}

\usepackage{ytableau}
\usepackage[enableskew]{youngtab} 

\usepackage{nicefrac}
\usepackage{xfrac}

\usepackage{color}

\usepackage{tikz}
\makeatletter
\newtheorem{rep@theorem}{\rep@title}
\newcommand{\newreptheorem}[2]{%
\newenvironment{rep#1}[1]{%
 \def\rep@title{#2 \ref{##1}}%
 \begin{rep@theorem}}%
 {\end{rep@theorem}}}
\makeatother

\newtheorem{theorem}{Theorem}[section]
\newreptheorem{theorem}{Theorem}
\newreptheorem{lemma}{Lemma}

\newcommand{\PP}{\mathbb{P}}
\newcommand{\FF}{\mathbb{F}}

\newcommand{\CC}{\mathbb{C}}
\newcommand{\ZZ}{\mathbb{Z}}

\def\phi{ \varphi }

\theoremstyle{definition}
\newtheorem{definition}[theorem]{Definition}

\theoremstyle{remark}
\newtheorem{remark}[theorem]{Remark}

\newcommand{\ccirc}[1]{\xymatrix@1{+<1ex>[o][F-]{#1}}}

\title[Testing Quantum Contextuality of Binary Symplectic Polar Spaces on a NISQ]{Testing Quantum Contextuality of Binary Symplectic Polar Spaces on a Noisy Intermediate Scale Quantum Computer}

\author{Fr\'ed\'eric Holweck}
\address{Université Bourgogne Franche-Comté, Laboratoire interdisciplinary Carnot de Bourgogne UMR6303, ICB-UTBM}
\email{frederic.holweck@utbm.fr}
\begin{document}

\maketitle

\begin{abstract}
The development of Noisy Intermediate Scale Quantum Computers (NISQC) provides for  the Quantum Information community new tools to perform quantum experiences from an individual laptop. It facilitates interdisciplinary research in the sense that theoretical descriptions of properties of quantum physics can be translated to experiments easily implementable on a NISCQ. In this note I test large state-independent inequalities for quantum contextuality on finite geometric structures encoding the commutation relations of the generalized $N$-qubit Pauli group. The bounds predicted by Non-Contextual Hidden Variables theories are strongly violated in all conducted  experiences.
\end{abstract}
\section{Introduction}
\begin{figure}[!h]
 \begin{center}
  \includegraphics[width=5cm]{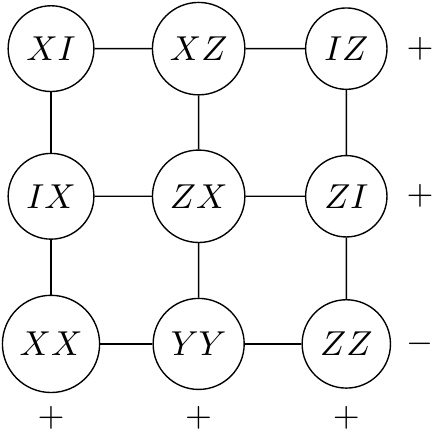}
  \caption{A Mermin-Peres magic square: Each node of the grid corresponds to a two-qubit Pauli observable with eigenvalues $\pm 1$. We use the shorthanded notation $A\otimes B\simeq AB$ where $A, B$ are Pauli matrices from $\{I, X,Y,Z\}$. Each line/row is a context, i.e. a set of mutually commuting operators. The product of three observables of a context (row/line) is $\pm I_4$ as indicated by the signs. Because there is an odd number of negative lines (here one) it is straightforward to check that there is no context-independent function $f$ which can assign $\pm 1$ to each node and satisfy also the sign constrains.}\label{fig:mermin}
 \end{center}
 \end{figure}

Quantum contextuality is a feature of quantum physics which contradicts the prediction of theories assuming realism and non-contextuality. Like entanglement, quantum contextuality was first considered as a paradox and is now recognized as a quantum resource for quantum information and quantum computation \cite{HWVE, BDBOR}. The Bell-Kochen-Specker Theorem \cite{B,KS}, often called Kochen-Specker Theorem (KS), is a no-go result that proves quantum contextuality by establishing that any Hidden-Variable (HV) theory that reproduces the outcomes of quantum physics should be contextual. 
In other words, if such an HV theory exists the deterministic functions describing the measurements are context-dependent, i.e. depend on the set of compatible, i.e. mutually commuting, measurements that are performed in the same experiment. It is this property of not being reproducible by any Non-Contextual Hidden Variables (NCHV) theory that we call quantum contextuality. In the 90's David Mermin \cite{M} and Asher Peres \cite{P} proposed  operator-based proofs of the Kochen-Specker Theorem using configurations of multi-qubits Pauli observables. Their most famous ``simple'' proof is the Magic Peres-Mermin square whose one example is reproduced in Figure \ref{fig:mermin}.
 
 Just a decade ago, testing quantum contextuality of the Mermin-Peres square in a lab was a great challenge \cite{ARBC}. But, in a recent preprint by Altay Dikme {et al.} \cite{DRLB}, it was shown  that measurements predicted by quantum physics for such a configuration of observables can be tested on an online NISQC\footnote{The Noisy Intermediate Scale Quantum Computers used in both this paper and \cite{DRLB} are the online accessible quantum computers provided by IBM through the IBM Quantum Experience \cite{ibmq}.}. The authors also checked that the outcomes of the measurements of their experiences cannot be explained by a NCHV theory. The Mermin-Peres grid has been investigated over the past 15 years from the perspective of finite geometry. It was, for instance proved \cite{HS}, that  Mermin's grids, as well as other configurations known as  Mermin's pentagrams, were the smallest, in terms of the number of contexts and operators, observable-based proofs of the KS Theorem. The geometry where those contextual configurations live is called the symplectic polar space of rank $N$ and order $2$ and will be denoted as $\mathcal{W}(2N-1,2)$. The correspondence between sets of mutually commuting $N$-qubit Pauli operators and totally isotropic subspaces of the symplectic polar space was established in \cite{SP,T,HOS}. This correspondence has been proved to be useful to describe quantum contextuality \cite{PS}, MUBs \cite{SPR}, symmetries of  black-hole entropy formulas \cite{LSVP} (as part of the black-hole/qubits correspondence) and other topics connecting quantum physics, error-correcting codes and space-time geometry \cite{LH}. It turns out that $\mathcal{W}(2N-1,2)$ is also useful to define robust quantum contextual inequalities \cite{C} and the purpose of this article is to test quantum contextuality on a NISQC based on the inequalities built from $\mathcal{W}(3,2)$ and $\mathcal{W}(5,2)$.

In Section \ref{sec:symplectic}, I recall the geometric construction of the symplectic polar space of rank $N$ and order $2$ that encodes the commutation relations of the generalized $N$-qubit Pauli group. 
In Section \ref{sec:ibm}, I present the results obtained on the IBM Quantum Experience when measuring  one-dimensional contexts of $\mathcal{W}(2N-1,2)$ for $N=2,3$. The results violate strongly the inequalities proposed by Ad\'an Cabello \cite{C} for testing quantum contextuality, i.e. detecting sets of measurements which contradict all NCHV theories. Finally, Section \ref{sec:conclusion} is dedicated to concluding remarks.

\section{The symplectic polar space of rank $N$ and order $2$}\label{sec:symplectic}
I recall in this section the correspondence between sets of mutually commuting $N$-qubit Pauli operators and totally isotropic subspaces of  the symplectic polar space of rank $N$ and order $2$, $\mathcal{W}(2N-1,2)$ \cite{SP,T,HOS}.
Let us denote by $\mathcal{P}_N\subset GL_{2^N}(\CC)$ the $N$-qubit Pauli group. Elements of $\mathcal{P}_N$ are operators $\mathcal{O}$ such that 
\begin{equation}\label{eq:operator}
 \mathcal{O}=sA_1\otimes A_2\dots\otimes A_N, \text{ with } s\in \{\pm 1,\pm i\} \text{ and } A_i\in \{I,X,Y,Z\},
\end{equation}
with $X,Y,Z$ being the usual Pauli matrices and $I$ the identity matrix.
\begin{equation}
 X=\begin{pmatrix}
    0 & 1\\
    1 & 0
   \end{pmatrix}, Y=\begin{pmatrix}
   0 & -i\\
   i & 0
   \end{pmatrix}, \text{ and } Z=\begin{pmatrix}
   1 & 0\\
   0 & -1
\end{pmatrix}.
\end{equation}

Like in Figure \ref{fig:mermin}, one will shorthand the notation of Eq. (\ref{eq:operator}) to $\mathcal{O}=sA_1A_2\dots A_N$. Also recall that the Pauli matrices $\{I,X,Y,Z\}$ can be expressed in terms of the matrix product of $Z$ and $X$ as:
\begin{equation}
 \begin{array}{cc}
  I=Z^0.X^0\leftrightarrow (0,0) & X=Z^0.X\leftrightarrow(0,1)\\
  Y=iZ^1.X^1 \leftrightarrow (1,1) & Z=Z^1.X^0 \leftrightarrow (1,0),
 \end{array}
\end{equation}
where $''.''$ is the usual matrix multiplication.
Thus Eq. (\ref{eq:operator}) can be expressed as:
\begin{equation}
 \mathcal{O}=s(Z^{\mu_1}.X^{\nu_1})(Z^{\mu_2}.X^{\nu_2})\dots(Z^{\mu_N}.X^{\nu_N}) \text{ with } s\in \{\pm 1,\pm i\}, \mu_i, \nu_j \in \{0,1\}.
\end{equation}
This leads to the following surjective map:
\begin{equation}\label{eq:map}
 \pi:\left\{\begin{array}{ccc}
      \mathcal{P}_N & \rightarrow & \FF_2^{2N}\\
      \mathcal{O}=s(Z^{\mu_1}.X^{\nu_1})(Z^{\mu_2}.X^{\nu_2})\dots(Z^{\mu_n}.X^{\nu_n}) & \mapsto& (\mu_1,\mu_2,\dots,\mu_N,\nu_1,\nu_2,\dots,\nu_N).
     \end{array}\right.
\end{equation}

The center of $\mathcal{P}_N$ is $C(\mathcal{P}_N)=\{\pm I_N,\pm i I_N\}$. Thus, $\mathcal{P}_N/C(\mathcal{P}_N)$ is an Abelian group isomorphic to the additive group $\FF_2^{2N}$, where $\FF_2=\{0,1\}$ is the two-elements field. Indeed, the surjective map given by Eq. (\ref{eq:map}) factors to the isomorphism $\mathcal{P}_N/C(\mathcal{P}_N)\simeq \FF_2^{2N}$.

Finally, considering the $(2N-1)$-dimensional projective space over $\FF_2$, $PG(2N-1,2)=\PP(\FF_2 ^N)$, one obtains a correspondence between non-trivial $N$-qubit observables, up to a phase $\{\pm 1,\pm i\}$, and points in the projective space $PG(2N-1,2)$.
\begin{equation}\label{eq:map2}
 \underline{\pi}:\left\{\begin{array}{ccc}
      \mathcal{P}_N/C(\mathcal{P}_N) & \rightarrow & PG(2N-1,2)\\
      \overline{\mathcal{O}}=s(Z^{\mu_1}.X^{\nu_1})(Z^{\mu_2}.X^{\nu_2})\dots(Z^{\mu_N}.X^{\nu_N}) & \mapsto& [\mu_1:\mu_2:\dots:\mu_N:\nu_1:\nu_2:\dots:\nu_N].
     \end{array}\right.
\end{equation}
The first correspondence does not say anything about commutation relations in $\mathcal{P}_N$. To recover these commutation relations, one needs to add an extra structure. Let us denote by $\mathcal{O}=s(Z^{\mu_1}.X^{\nu_1})(Z^\mu_2.X^\nu_2)\dots(Z^{\mu_N}.X^{\nu_N})$ and $\mathcal{O}'=s'(Z^{\mu_1'}.X^{\nu_1'})(Z^{\mu_2'}.X^{\nu_2'})\dots(Z^{\mu_N'}.X^{\nu_N'})$ two representatives of the classes $\overline{\mathcal{O}}$ and $\overline{\mathcal{O}'}$. The classes will commute iff $\mathcal{O}.\mathcal{O}'=\mathcal{O}'.\mathcal{O}$. But a straightforward calculation shows that 
\begin{equation}\mathcal{O}.\mathcal{O}'=ss'(-1)^{\sum_i \nu_i\mu_i'}(Z^{\mu_1+\mu_1'}.X^{\nu_1+\nu_1'})(Z^{\mu_2+\mu_2'}.X^{\nu_2+\nu_2'})\dots (Z^{\mu_N+\mu_N'}.X^{\nu_N+\nu_N'})\end{equation}
and,
\begin{equation}\mathcal{O}'.\mathcal{O}=ss'(-1)^{\sum_i \nu_i'\mu_i}(Z^{\mu_1+\mu_1'}.X^{\nu_1+\nu_1'})(Z^{\mu_2+\mu_2'}.X^{\nu_2+\nu_2'})\dots (Z^{\mu_N+\mu_N'}.X^{\nu_N+\nu_N'}).\end{equation}
Therefore the classes $\mathcal{O}$ and $\mathcal{O}'$ commute iff $\sum_{i=1} ^N \mu_i\nu_{i}'+\mu_i'\nu_i=0$. 

Let us define on $PG(2N-1,2)$ the sympletic form:
\begin{equation}
 \langle p,q\rangle=\sum_{i=1} ^N p_iq_{N+i}+p_{N+i}q_i,
\end{equation}
with $p=[p_1:\dots:p_{2N}]$ and $q=[q_1:\dots:q_N]$.
We can now state  the definition of the symplectic polar space of rank $N$ and order $2$:.
\begin{definition}
 The space of totally isotropic subspaces\footnote{A totally isotropic subspace is a linear space of $PG(2N-1,2)$ on which the symplectic form vanishes identically.} of $PG(2N-1,2)$ for a nondegenerate symplectic form $\langle,\rangle$ is called the symplectic polar space of rank $N$ and order $2$ and is denoted by $\mathcal{W}(2N-1,2)$.
\end{definition}
From the previous discussion it is clear that points of $\mathcal{W}(2N-1,2)$ are in correspondence with non-trivial $N$-qubit Pauli operators and linear subspaces of $\mathcal{W}(2N-1,2)$ are sets of mutually commuting operators in $\mathcal{P}_N$ (i.e. contexts). For small values of $N$ the geometry of $\mathcal{W}(2N-1,2)$ has been studied in detail in \cite{SPPH, LHS}. I recall some of their distinguished geometric features and their relations with quantum contextuality.
\subsection{The doily, $\mathcal{W}(3,2)$, and its Mermin-Peres grids}
\begin{figure}[!h]
 \includegraphics[width=6cm]{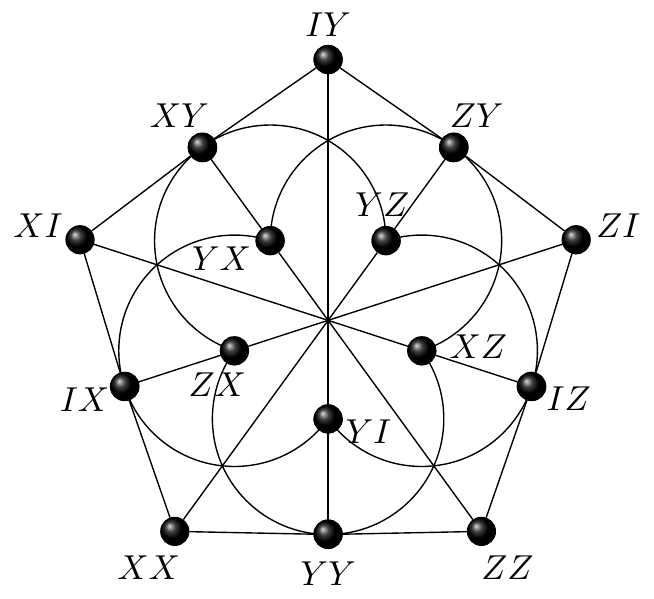}
 \caption{The symplectic polar space $\mathcal{W}(3,2)$, aka the doily.
 As a point-line geometry, the doily is a $15_3$-configuration, $15$ lines, $15$ points, $3$ points per lines and $3$ lines through each point. Out of 245 342 non-isomorphic $15_3$-configurations, $\mathcal{W}(3,2)$ is the only one that is triangle free.}\label{fig:doily}
\end{figure}

For $N=2$ the symplectic polar space $\mathcal{W}(3,2)$ comprises $15$ points (two-qubit non-trivial observables) and $15$ lines (contexts). 
 The structure of $\mathcal{W}(3,2)$, with its points labelled by canonical ($s=+1$) representatives of $\mathcal{P}_2$ is illustrated in Figure \ref{fig:doily}. 

\begin{figure}[!h]
 \includegraphics[width=6cm]{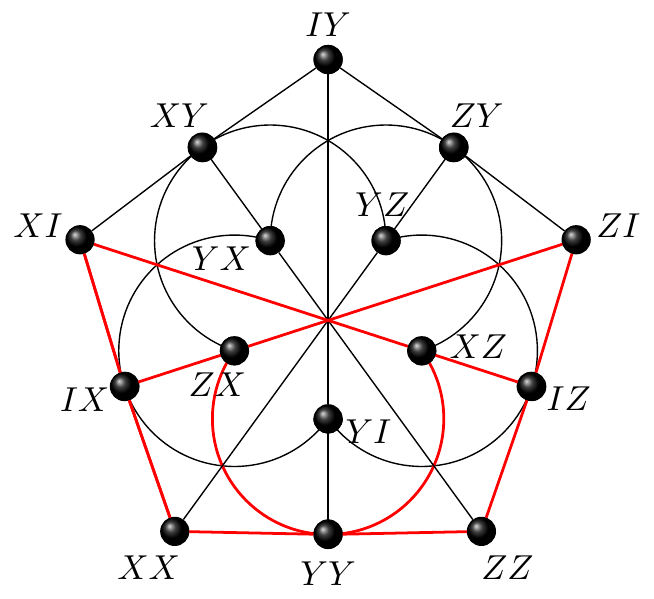}
 \caption{A Mermin-Peres magic square sitting in $\mathcal{W}(3,2)$: 9 more grids can be obtained from this canonical labeling. The fact that $\mathcal{W}(3,2)$ is triangle-free implies that all grids  furnish an operator-based proof of KS Theorem \cite{HS}.}\label{fig:doilymermin}
\end{figure}

Mermin-Peres magic squares live inside the doily as subgeometries \cite{SPPH}. More precisely they are geometric hyperplanes of the doily, i.e. subsets such that any line of the geometry is either fully contained in the subset, or intersects the subset at only one point. The Mermin-Peres magic square of Figure \ref{fig:mermin}, embedded in $\mathcal{W}(3,2)$, is reproduced in Figure \ref{fig:doilymermin}.

An alternative way of defining the Mermin-Peres squares as subgeometries of $\mathcal{W}(3,2)$ is to consider  hyperbolic quadrics. Consider a non-degenerate quadratic form defined by 
\begin{equation}\label{eq:q0}
 \mathcal{Q}_0(x)=x_1x_3+x_2x_4, \text{ with }  x=[x_1:x_2:x_3:x_4]\in \mathcal{W}(3,2).
\end{equation}

Such a quadric is said to be hyperbolic, \cite{VL,LHS}, and the zero locus of $\mathcal{Q}_0$, to be denoted by $\mathcal{Q}_0^{+}(3,2)$, comprises all $9$ symmetric two-qubit observables, i.e. all operators with an even number of $Y$'s. In other words  $\mathcal{Q}_0^{+}(3,2)$ is the Mermin-Peres square of Figure \ref{fig:mermin} and \ref{fig:doilymermin}. To obtain the remaining nine Mermin-Peres squares, one can look at the following hyperbolic quadrics:
\begin{equation}\label{eq:qp}
 \mathcal{Q}_p^{+}(3,2)=\{x\in \mathcal{W}(3,2), \mathcal{Q}_p(x):= \mathcal{Q}_0(x)+\langle p,x\rangle=0, \text{ with } \mathcal{Q}_0(p)=0\}.
\end{equation}
In terms of operators, $\mathcal{Q}_p^{+}(3,2)$ accommodates those two-qubit observables that are either symmetric and commute with $\mathcal{O}_p$ or skew-symmetric and anti-commute with $\mathcal{O}_p$. 

\subsection{$\mathcal{W}(5,2)$ and its hyperbolic quadrics}
\begin{figure}[!h]
 \includegraphics[width=6cm]{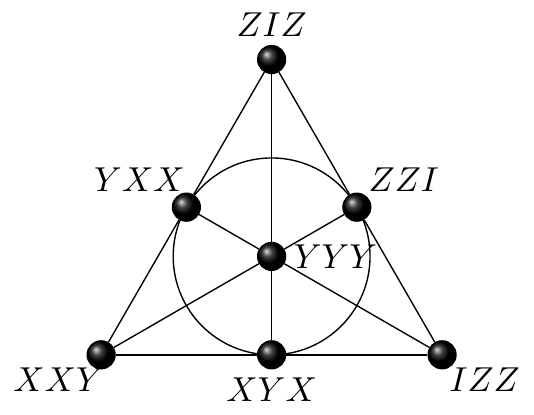}
 \caption{An example of Fano plane in $\mathcal{W}(5,2)$. This Fano plane is a negative context in the sense that the product of all opertators gives $-I_{8}$. It contains seven one-dimensional linear contexts (the lines). The three lines meeting at $YYY$ are negative.}\label{fig:fano}
\end{figure}

For $N=3$ the symplectic polar space contains $63$ points (non-trival three-qubit observables), $315$ lines ($1$-dimensional linear contexts) and $135$ Fano plane ($2$-dimensional linear contexts). An example of a three-qubit Fano plane of $\mathcal{W}(5,2)$ is reproduced in Figure \ref{fig:fano}. Subgeometries of $\mathcal{W}(5,2)$ have been studied in \cite{LHS} and a full description of the hyperplanes of $\mathcal{W}(2N-1,2)$ has been carried out in \cite{VL}. 

Hyperbolic quadrics, for instance, form one class of hyperplanes in $\mathcal{W}(5,2)$ and their definition is a straightforward generalization\footnote{The generalization is in fact, also straightforward for the general  $N$, see \cite{VL}.} of Eq. (\ref{eq:q0}) and Eq. (\ref{eq:qp}). The quadric corresponding to the symmetric three-qubit observables is
\begin{equation}
 \mathcal{Q}_0^{+}(5,2)=\{x\in \mathcal{W}(5,2), \mathcal{Q}_0(x):=x_1x_4+x_2x_5+x_3x_6=0\}.
\end{equation}
$35$ additional quadrics can be defined as,
\begin{equation}
 \mathcal{Q}_p^{+}(5,2)=\{x\in \mathcal{W}(5,2), \mathcal{Q}_p(x):=\mathcal{Q}_0(x)+\langle p,x\rangle=0, \text{ with } \mathcal{Q}_0(p)=0\}.
\end{equation}

A hyperbolic quadric in $\mathcal{W}(5,2)$ consists of $35$ points, $105$ lines and $15$ Fano planes. A combinatorial description of the hyperbolic quadrics is provided in \cite{SZ}. It is, for example, known that the $35$ points of $\mathcal{Q}_p^+(5,2)$ splits into $15+20$ where the first set forms a doily inside $\mathcal{Q}_p^+(5,2)$ while the $20$ off-doily points make ten complementary pairs. Through each point of the the pair pass nine lines and those nine lines intersect the doily in a grid (same grid for two points of the same pair). One, therefore, has a partition of the $105$ lines of a hyperbolic quadric into $90=9\times 10$ off-doily lines (the $9$ lines from the $10$ pairs of points) plus $15$ lines of the doily. Those geometric properties will be useful in the next section to discuss the conditions that can be satisfied by a NCHV theory for the configuration given by $\mathcal{Q}_0 ^{+}(5,2)$.

\begin{remark}
Small contextual configurations can be found in $\mathcal{W}(5,2)$ including, as already mentioned, copies of Mermin-Peres squares  but also Mermin's pentagrams (configurations featuring $10$ observables and $5$ contexts). It has been proved that there are $12 096$ distinguished Mermin's pentagrams in $\mathcal{W}(5,2)$ \cite{PSH,LS}. This number is remarkably equal ot the order of  the automorphism group of the smallest Split Cayley hexagon, a notable $63_3$ configuration that can be embedded into $\mathcal{W}(5,2)$ (see the conclusion and \cite{PSH,LSVP} for more details).
\end{remark}

\section{Measuring the contextuality of $\mathcal{W}(2N-1,2)$ with the IBM Quantum Experience}\label{sec:ibm}
\subsection{Macroscopic state-independent inequalities}
Let us consider  a subgeometry $\mathcal{G}$ of $\mathcal{W}(2N-1,2)$, i.e. a set of points ($N$-qubit observables) and lines (contexts made of three observables). The inequalities, built from $\mathcal{G}$, that we tested on the IBM Quantum Experience are of the following form:
\begin{equation}\label{eq:rio1}
 \chi_\mathcal{G}=\sum_{i=1} ^S \langle \mathcal{C}_i\rangle-\sum_{i=S+1} ^M \langle \mathcal{C}'_i\rangle \leq b_\mathcal{G},
\end{equation}
where $\langle\mathcal{C}_i\rangle$ denotes the mean value of the product of three observables measured sequentially on the same positive context (i.e. a positive line of $\mathcal{G}$) while $\langle\mathcal{C}'_i\rangle$ denotes the mean value of the product of the three observables measured sequentially on a negative context (i.e. a negative line) of $\mathcal{G}$. The configuration is, therefore, made of $M$-contexts with $S$ positive ones and $M-S$ negative ones.

The upper bound $b_\mathcal{G}$ takes different values if we assume that the results of the measurements can be explained by a NCHV theory or if we consider the prediction of Quantum Mechanics (QM).
More precisely, we have, as shown in \cite{C}:
\begin{equation}\label{eq:rio2}
 b_{\mathcal{G}}^{NCHV}= 2P-M\ \ \ \ b_\mathcal{G}^{QM}= M,
\end{equation}
where $P$ is the maximum number of predictions that can be satisfied by a NCHV theory. Clearly $P\geq S$ as  it is always possible to assign $+1$ to all possible observables as a predefinite value and all positive constrains will be satisfied.

The Mermin-Peres square configurations, $\mathcal{Q}_{p}^{+}(3,2)$, can be used to provide symple examples of inequalities of type given by Eq. (\ref{eq:rio1}). In this case one has  $b_{\mathcal{Q}_p^{+}(3,2)}^{NCHV}=4$ and $b_{{\mathcal{Q}_p^{+}(3,2)}}^{QM}=6$. 
 In \cite{DRLB} it is $\chi_{\mathcal{Q}_{p}^{+}(3,2)}$ for $p=XX$ that was tested on the IBM Quantum Experience and the comparison that the authors make of their experimental results to a general mixture of results obtained from a NCHV theory is equivalent to comparing their measured value of $\chi^{\text{exp}}_{\mathcal{Q}_p^{+}(3,2)}$ with $b_{\mathcal{Q}_p^{+}(3,2)}^{NCHV}$. 

Ad\'an Cabello studied in \cite{C} the robustness of Eq. (\ref{eq:rio1}) in order to find inequalities that would overperform the ones produced by  Mermin-Peres squares. He proved by introducing  the tolerated error per correlation, $\varepsilon=\dfrac{b_\mathcal{G} ^{QM}-b_\mathcal{G} ^{NCHV}}{M}$, that the most robust inequalities are given by considering all possible contexts made of three observables in $\mathcal{P}_N$. In this case one has $P=S$ and in our geometric language Cabello's result can be rephrased by saying that the best  inequalities of type Eq. (\ref{eq:rio1}) to test quantum-contextuality are  $\chi_{\mathcal{W}(2N-1,2)}$. These are the inequalities that we tested on the IBM Quantum Experience.
\subsection{Results of the experiments}
The IBM Quantum Experience is an online platform launched by IBM in 2016 which gives access to NISQC from $5$ to $16$ qubits \cite{ibmq}. A graphical interface allows the user to easily generate quantum circuits and run them on different backends\footnote{The different quantum computers available on the IBM Quantum Experience have names of type 'ibmq\_athens', 'ibmq\_vigo', 'ibmq\_santiago'. The machines differ from each other by their connectivity architecture and robustness of gates, see \cite{ibmq}.}. An open-source  software development kit, Qiskit \cite{qiskit}, is also available to create and run programs on the machines of the IBM Quantum Experience. Due to a large number of measurements to perform, I used Qiskit to generate all possible measurements and send them to the IBM Quantum Experience.
All the Qiskit codes and tables of the numerical results obtained are availble at \url{https://quantcert.github.io/Testing_contextuality}.
Since 2016 several researchers have been using the IBM Quantum Experience as an experimental platform to launch quantum computations and the reliability of the machines has constantly increased since then \cite{A,GM,Lee,Li,SSBP,H,RBBP,DRLB}. 
\subsection{Measuring a context}
\begin{figure}[!h]
\[
\Qcircuit @C=1em @R=.7em {
\lstick{\ket{q_0}} &\gate{H} & \ctrl{3} & \gate{H} & \qw & \qw & \qw & \qw & \qw\\
\lstick{\ket{q_1}} &\qw & \qw & \gate{S^\dagger} & \gate{H} & \ctrl{2} & \gate{H} & \gate{S} &\qw\\
\lstick{\ket{q_2}} &\qw & \qw & \qw & \qw & \qw & \ctrl{1} & \qw &\qw \\
\lstick{\ket{q_3}=\ket{0}} &\qw & \targ & \qw & \qw & \targ & \targ & \qw& \meter }
\]
\caption{Quantum circuit measuring the observable $XYZ$ on the auxiliary qubit $\# 3$. The rotation (resp. $H$ and $S^\dagger T$) prior to the $CNOT$ gates corresponds to a change of basis (resp. $X$ and $Y$). In order to anticipate the sequential measurement that will follow, the counter-rotation is applied after the $CNOT$ gates.}\label{fig:XYZ}
\end{figure}
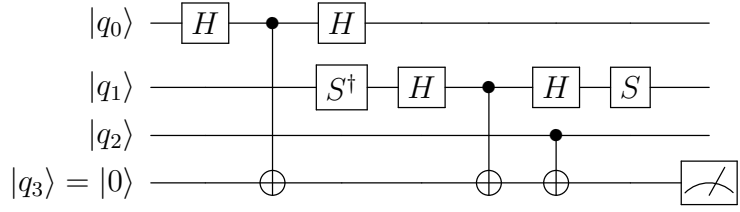

We follow the strategy of \cite{DRLB}, save for  a few variations explicitly indicated in the text. The first two constrains are the following:
\begin{itemize} 
 \item the measurement performed for each observable should be nondestructive in order to get a sequential measurement. This can be achieved by introducing an auxiliary qubit;
 \item the IBM Quantum Experience only allows measurement in the $Z$-basis. Thus rotation prior to the measurement in the $X$ and $Y$ basis should be made.
\end{itemize}
Figure \ref{fig:XYZ} illustrates how to measure on the IBM Quantum Experience the observable $XYZ$ on an auxiliary qubit. The outcome measured on qubit $\#3$ is the product of the observed eigenvalues of the three observables.

The sequential measurements of three observables on one context will be obtained by the concatenation of three circuits similar to Figure \ref{fig:XYZ}, like in \cite{DRLB}. However, contrary to the circuits generated in \cite{DRLB}, I will:
\begin{itemize}
 \item not make any circuit optimization before sending the calculation to the IBM Quantum Experience. The main reason being that for the case $N=3$, I had to measure  $315$ contexts and could not optimize each circuit one by one;
 \item add barriers between each observable measurement to force the IBM Quantum Experience to not optimize and simplify the calculation. Indeed once the calculation is sent to the IBM Quantum Experience, a {\em transpilation} step is performed to transform the initial circuit to a circuit which respects the connectivity of the actual quantum machine\footnote{The different quantum machines have different topology in terms  of connectivity. This means that two-qubit gates, like $CNOT$, cannot be always directly  implemented between two qubits as indicated in the program. The transpilation process translates the initial program to an equivalent circuit, which respects the machine connectivity \cite{ibmq}.}. This transpilation phase may sometimes simplify  the computation too much. The addition of barriers forces the machine to perform the three sequential measurements one after the other.
\end{itemize}

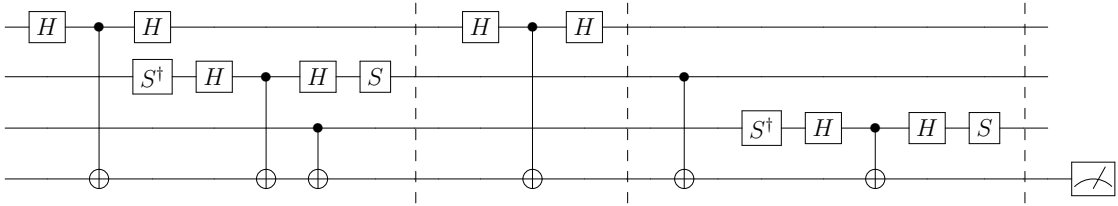
\begin{figure}[!h]
\begin{center}
\[\resizebox{.9\linewidth}{!}{
\Qcircuit @C=1em @R=.7em 
{&\gate{H} & \ctrl{3} & \gate{H} & \qw & \qw & \qw & \qw&\qw\barrier{3} & \qw & \gate{H} & \ctrl{3} & \gate{H} &\qw\barrier{3} & \qw &\qw & \qw & \qw &\qw &\qw &\qw&\qw &\qw\barrier{3} &\qw  \\
 &\qw & \qw & \gate{S^\dagger} & \gate{H} & \ctrl{2} & \gate{H} & \gate{S} &\qw & \qw &\qw & \qw &\qw & \qw & \qw & \ctrl{2} & \qw & \qw &\qw &\qw &\qw& \qw &\qw &\qw  \\
&\qw & \qw & \qw & \qw & \qw & \ctrl{1} & \qw &\qw &\qw&\qw & \qw &\qw & \qw & \qw & \qw &\qw & \gate{S^\dagger} & \gate{H} & \ctrl{1} & \gate{H} & \gate{S} &\qw &\qw\\
&\qw & \targ & \qw & \qw & \targ & \targ & \qw& \qw &\qw&\qw & \targ &\qw & \qw & \qw &\targ &\qw &\qw& \qw &\targ & \qw & \qw & \qw &\qw& \meter}}
\]
\end{center}
\caption{Sequential measurement of $XYZ-XII-IZY$: The product of the three eigenvalues corresponding to the three measurements is obtained on the auxiliary qubit $\#4$.}\label{fig:contextmeasurement}
\end{figure}

Figure \ref{fig:contextmeasurement} shows the sequential measurement of a typical three-qubit context as it is sent to the IBM Quantum Experience by our program.
Both new constrains make the calculation more fragile as it increases the noise by adding non necessary gates (like the two Hadamard gates on the first qubit of Figure \ref{fig:contextmeasurement}). However, due to the robustness of $\chi_{\mathcal{W}(3,2)}$ and $\chi_{\mathcal{W}(5,2)}$, these contrains will not be an obstacle to test quantum contextuality.

%
\subsection{Results for $\chi_{\mathcal{W}(3,2)}$}
I first  tested the inequality corresponding to the doily. Our program consists of two main parts. One script generates all points and lines\footnote{with signs given by the canonical labeling by two-qubit observables.} of $\mathcal{W}(3,2)$ following the prescriptions of Section \ref{sec:symplectic} and the second part  generates with Qiskit all circuits corresponding to each context to be measured, then the calculation is sent to the IBM Quantum Experience, which sends back the results of the different measurements. In order to obtain significant statistics, each measurement is performed $8192$, times which is the maximum number of shots allowed by the IBM Quantum Experience. All scripts are available at \url{https://quantcert.github.io/Testing_contextuality}.
The numerical outcomes are reproduced in Table \ref{tab:doily} for an experiment conducted on January 4, 2021, on the backend ibmq\_casablanca\footnote{We conducted our experiences between November 17, 2020, and January 5, 2021, on different backends: ibmq\_valencia, ibmq\_santiago, ibmq\_yortown. We always obtained violation of the NCHV inequalities except with the backend ibmq\_rome. The backend ibmq\_casablanca, accessible with an ibm-q-research account, provided the best results.}. Each context $\mathcal{C}_i$ is considered as a dichotomic experiment with two outcomes $\{+1,-1\}$, the standard deviation is therefore calculated as $\sigma_{\mathcal{C}_i}=\sqrt{\dfrac{p(1-p)}{n_\text{exp}}}$ where $p$ is the probability of measuring $+1$ and $n_{\text{exp}}$ is the number of measurements (experiences) which are preformed on the context $\mathcal{C}_i$.
Based on the numerical values of Table \ref{tab:doily}, one gets
\begin{equation}
 \chi_{\mathcal{W}(3,2)} ^{\text{exp}}=12.27
\end{equation}
with total standard deviation $\sigma_\chi=0.012$. Therefore, the classical bound $b^{NCHV}=9$ is violated by more than $270$ standard deviations, i.e. more than $10$ times of the violation obtained for $\mathcal{G}=\mathcal{Q}_p^{+}(3,2)$ as given in \cite{DRLB}.
\begin{table}[!h]
 \begin{tabular}{|c|c|c|c|c|}
 \hline
  Context $\mathcal{C}_i$ & $\# 1$ & $\# -1$ & $\langle \mathcal{C}_i\rangle$ & std \\
  \hline
$IX-XI-XX$ & $7337$ & $855$& $0,7913$ & $0,0034$\\
$IX-YI-YX$  & $7283$ & $909$ & $0,7781$ & $0,0035$\\
$IX-ZI-ZX$& $7213$ & $979$ &$0,7610$ &$0,0036$\\
$IY-XI-XY$& $7511$ & $681$ & $0,8337$ & $0,0031$\\
$IY-YI-YY$& $7465$ &$727$ & $0,8225$  &$0,0031$\\
$IY-ZI-ZY$& $7275$ &$917$ & $0,7761$ & $0,0035$\\
$IZ-XI-XZ$& $7644$ &$548$ & $0,8662$&$0,0035$\\
$IZ-YI-YZ$& $7713$ &$479$ &$0,8831$ & $0,0028$\\
$IZ-ZI-ZZ$& $7822$ &$370$ & $0,9097$& $0,0026$\\
$XX-YY-ZZ$& $735$& $7457$&$-0,8206$& $0,0023$\\
$XX-YZ-ZY$& $7567$& $625$ &$0,8474$ &$0,0032$\\
$XY-YX-ZZ$& $7552$ &$640$&$0,8437$ &$0,0029$\\
$XY-YZ-ZX$& $968$ &$7224$ & $-0,7637$&$0,0030$\\
$XZ-YX-ZY$& $1054$ &$7138$ & $-0,7427$& $0,0036$\\
$XZ-YY-ZX$& $7499$ & $693$ & $0,8308$& $0,0037$\\
  \hline
 \end{tabular}
 
\caption{Numerical values obtained to measure $\chi_{\mathcal{W}(3,2)}$ on January $4$, $2021$ on the IBM Quantum Experience, ibmq\_casablanca. Each measurement was repeated $8192$ times.}\label{tab:doily}
\end{table}

\subsection{Results for $\chi_{\mathcal{W}(5,2)}$}
As anticipated in \cite{C}, the violation is even more pronounced for the $N=3$-qubit symplectic polar space  $\mathcal{W}(5,2)$. The calculation involves the measurement of $315$ contexts, which are generated and measured by two Python-Qiskit scripts, like those for the $\mathcal{W}(3,2)$ case.

In January $4$, I obtained on the ibmq\_casablanca backend the following value:
\begin{equation}
 \chi_{\mathcal{W}(5,2)}^{\text{exp}}=236.57
\end{equation}
with standard deviation $\sigma_{\chi_{\mathcal{W}(5,2)}^{\text{exp}}}=0.064$. Thus the NCHV bound, $b^{NCHV}=135$, is violated by more than $1500$ standard deviations.

I also calculated $\chi_{\mathcal{Q}_0^{+}(5,2)}$ because, as explained in Section \ref{sec:symplectic}, hyperbolic quadrics are natural generalization of  Mermin-Peres grids for all $N$. To obtain the upper bound $b_{\mathcal{Q}_0^{+}(5,2)}^{NCHV}$, one needs to calculate the maximum number of predictions that can be achieved with a NCHV theory. Using the canonical labelling, a hyperbolic quadric contains either $78$ (resp. $27$)  or $66$ (resp. $39$) positive contexts (resp. negatives ones). It is thus clear  that $P\geq 78$. 
Now the $105=90+15$ splitting of the lines of $\mathcal{Q}_0^{+}(5,2)$, where all $90$ off-doily lines can be partitioned into $10$ groups of nine intersecting the doily in its $10$ Mermin-Peres grids, shows that any NHCV theory that could satisfy the prediction of one of the $27$ negative lines would make it possible to have a NHCV theory satisfying more constrains for some of the $10$ grids. But this is not possible and therefore one may conclude that for hyperbolic quadrics $P=78$ and thus $b_{\mathcal{Q}_0^{+}(5,2)}^{NCHV}=51$.
In January 4, 2021, I obtained the following experimental value on the backend ibmq\_casablanca:
\begin{equation}
 \chi_{\mathcal{Q}_0^{+}(5,2)}^{\text{exp}}=82.17,
\end{equation}

\noindent with the standard deviation $\sigma=0.035$, corresponding to a violation of the NCHV bound by $890$ standard deviations. Detail of our results for $\mathcal{W}(5,2)$ and $\mathcal{Q}_0^+(5,2)$ are available on \url{https://quantcert.github.io/Testing_contextuality}.

\section{Conclusion}\label{sec:conclusion}
In this note, I tested  on the IBM Quantum Experience the  inequalities of \cite{C} to detect quantum contextuality based on configurations of two-qubit and three-qubit observables. The measurements performed follow the experiments of \cite{DRLB} where the famous Mermin-Peres square configuration was successfully tested. The results show that, despite the current imperfections of the IBM quantum machines, the bounds predicted by NCHV theories are strongly violated. Our first calculation, regarding a configuration of two-qubit observables, involves the measurement of $15$ different contexts and the second experiment, in the three-qubit case, entails $315$ contexts. One also used our programs to test subgeometries in the three-qubit case.

Interestingly, the easy use of online available NISCQ opens up new paths for both research and scientific education. As I tried to emphasize in this note, the studies of quantum contextuality from finite geometric configurations \cite{HS,SP,T,HOS,PS,LS} can now translate to experiments on real quantum machines. In particular, it would be very exciting to see the symmetries of some configurations manifesting in the results of some quantum experiments. For instance in the three-qubit case, the Split Cayley hexagon of order $2$ is a configuration made of the $63$ observables of $\mathcal{W}(5,2)$ and only $63$ contexts which form a generalized hexagon \cite{LSVP}. The automorphism group of this configuration is  $\text{SL}(7,2)$, the quotient of the Weyl group $W(E_7)$ by $\ZZ_2$, and the connection with the Mermin-pentagrams contextual configurations was studied from the geometrical perspective. Now, it is possible to check the contextuality of the Split Cayley hexagon by the same techniques as developed here. I believe this is not only scientifically interesting, but it also emphasizes the strong interdisciplinary dimension of quantum information, which is also very valuable from a training perspective. 
\section*{Acknowledgment}
This work was supported by the French Investissements d'Avenir programme, project ISITE-BFC (contract ANR-15-IDEX-03). I acknowledge the use of the IBM Quantum Experience and the IBMQ-research program. The views expressed are those of the author and do not reflect the official policy or position of IBM or the IBM Quantum Experience team. One would like to thank the developers of the open-source framework Qiskit as well as Metod Saniga for his comments on an earlier version of the paper and Ad\'an Cabello for explaining us the ``Rio Negro'' inequalities during his stay in Besan\c{c}on for the IQUINS meeting in 2017.

\end{document}